\documentclass{emulateapj}
\usepackage{psfig}

\shorttitle{KH15D}
\shortauthors{Agol et al.}

\begin{document}


\title{Spectropolarimetry and Modeling of the Eclipsing T Tauri Star KH 15D}

\author{Eric Agol\altaffilmark{1,2,3}, Aaron J. Barth\altaffilmark{4,5},  Sebastian Wolf\altaffilmark{4},
David Charbonneau\altaffilmark{4}}

\altaffiltext{1}{Theoretical Astrophysics, MS 130-33, California Institute of Technology, Pasadena, CA 91125 USA; agol AT tapir.caltech.edu}
\altaffiltext{2}{Chandra Fellow}
\altaffiltext{3}{Department of Astronomy, University of Washington, Box 351580,
Seattle, WA 98105; agol AT astro.washington.edu}
\altaffiltext{4}{Department of Astronomy, MS 105-24, California Institute of Technology, Pasadena, CA 91125}
\altaffiltext{5}{Hubble Fellow}
\slugcomment{}


\begin{abstract}
KH 15D is a strongly variable T Tauri star
in the young star cluster NGC 2264 that shows a decrease in flux of 3.5
magnitudes lasting for 18 days and repeating every 48 days.
The eclipsing material is likely due to orbiting dust or rocky bodies 
in a partial ring or warped disk that periodically occults the star.
We measured the polarized 
spectrum in and out of eclipse at the Keck and Palomar observatories.
Outside of the eclipse, the star exhibited low polarization 
consistent with zero.  During eclipse, the polarization increased dramatically 
to $\sim 2$\% across the optical spectrum, while the spectrum had the same 
continuum shape as outside of eclipse and exhibited emission lines of much larger 
equivalent width, as previously seen.  From the data, we conclude that 
(a) the scattering region is uneclipsed;  (b) the scattering is 
nearly achromatic;  (c) the star is likely completely eclipsed
so that the flux during eclipse is entirely due to scattered light,
a conclusion also argued for by the shape of the ingress and egress.
We argue that the scattering is not due to electrons, but may be due
to large dust grains of size $\sim 10\mu$m, similar to the interplanetary 
grains which 
scatter the zodiacal light.  We construct a warped-disk model with an extended
dusty atmosphere which reproduces the main features of the lightcurve, namely 
(a) a gradual decrease before ingress due to extinction in the atmosphere (similar 
for egress);  (b) a sharper decrease within ingress due to the optically-thick 
base of the atmosphere; (c) a polarized flux during eclipse which is 
0.1\% of the total flux outside of eclipse, which requires no fine-tuning
of the model.  The inclination of the warp is set by the duration of ingress
and egress assuming that the warp is located at the Keplerian radius, and 
the inclination of the observer is then determined by the duration of the eclipse.

\end{abstract}



\section{Introduction}

T Tauri stars are thought to be young stars that are still accreting 
gas through a disk.  \citet{kea98} discovered T Tauri star
KH 15D (K7 V, $d=760$ pc), which shows an eclipse for 1/3 of its period, 
implying that it is obscured by faint circumstellar material rather than 
a stellar companion;  earlier data showed the star rebrighten at mid-eclipse, 
although this has weakened in recent data \citep{ham01}.  This 
indicates that the obscuring material subtends a large angle, which may
herald the presence a low mass stellar companion to shepherd the material 
via resonant interaction or to create a distorted disk 
\citep{bry00}.  In addition, the duration 
of the eclipse trough is lengthening over time, changing from 16 days 
to 18 days over 5 years, and data from 50-90 years ago showed no evidence 
for eclipses \citep{win03}, which means that the obscuring material
is rapidly evolving.  During eclipse, residual flux is still present
at a level of $\sim$1/20 of the flux outside of eclipse indicating 
that either (a) the obscuring material is porous (covering 
factor is $<1$), (b) the eclipse is total, but scattered 
light ``fills in'' the eclipse, creating some residual flux, (c) some
flux is scattered and some transmitted.  The scattering interpretation
may be favored as the star becomes slightly bluer during eclipse as one 
would expect from scattered light, although the fluctuations in color
are comparable to the color difference \citep{her03}.  In addition,
the equivalent width of H$\alpha$ increases dramatically from 
2 \AA\  to $\sim 30$\AA\ indicating a change from an
unobscured ``weak-line'' T Tauri star to a classical T Tauri
star!

The eclipse provides a natural ``coronagraph'' during which the
suppression of stellar light allows any
scattering by surrounding material to become more prominent.  We
attempted to distinguish the above possibilities by carrying out 
spectropolarimetry inside and outside of the eclipse.
Since the unpolarized flux of the star dilutes the percentage 
of polarization, we expected that the percentage polarization due 
to scattered light would increase by a factor of at least 20 during 
the eclipse if the scattering material lies exterior to the occulting 
material and if the starlight is intrinsically unpolarized.  This scenario is analogous
to Type 1 active galactic nuclei (AGN) which are unobscured and show low 
polarization versus Type 2 AGN which are obscured and show much higher 
polarization \citep{ant93}.  Some T Tauri stars show an 
anti-correlation between total flux and polarization \citep[e.g.][]{mek99,
men03}, which may be exaggerated in KH 15D as the flux variations are 
much more extreme.  On the other hand, if the residual flux 
is only due to incomplete coverage by the occulting material, then 
the polarization will not change.  Interstellar foreground polarization,
which is measured to be 0.1--0.4\% in the V-band for NGC 2264 \citep{bre72}, 
will not change during eclipse.  
Before our observations, the optical polarization of KH 15D had not yet 
been measured, but during eclipse it shows the characteristics of 
classical T Tauri stars which have a mean polarization of 1.2\% with
a standard deviation of 0.8\% and maximum polarization of 2.5\% \citep{men91}.

In \S 2 we summarize the spectropolarimetric observations, 
in \S 3 we give the results of the observations, in \S 4 we discuss
the implications of these results for models of this T Tauri star, and
in \S 5 we review our conclusions.

\section{Observations}

The primary observational goal was to measure the difference in polarization 
between eclipse and non-eclipse.  Outside eclipse, the star is sufficiently
bright to observe at the Palomar 200-inch telescope ($V \sim 16$), but 
during the eclipse the star
is a factor of $\sim 20$ fainter which required using Keck to
obtain a similar number of photons in a reasonable observing time.  The Palomar 
observations were taken with the Double Spectrograph spectropolarimeter
\citep{oke82, goo91}, while the Keck I observations were taken with the Low 
Resolution Imaging Spectrograph polarimeter \citep{oke95,goo91}.  Both
instruments add dual-beam polarimetry optics
to a double spectrograph incorporating dichroics to
separate the blue and red beams.  The observations and reductions were
done using standard spectropolarimetry techniques \citep{mil88}. The
observations of KH 15D were performed in sequences of four 15-minute 
exposures at four different positions of a rotating half-wave plate.
Each four-exposure sequence allows measurement of the Stokes parameters 
$q$ and $u$.  For each hour of observations, the position angle of
the spectrograph slit was aligned with the parallactic angle for the
midpoint of the hour.
The instrumental setup is described in Table 1, while the observational
summary is given in Table 2.   Null, polarized, and flux standards were
observed during each night.  Since instrumental polarization can be
present in the Palomar 200-inch \citep{ogl99}, we observed four null 
standards each at two perpendicular slit position angles.
We used the measured polarization of the null standards as an estimate
of the mean and uncertainty of the instrumental polarization, and
we subtracted the mean instrumental Stokes parameters $q$ and $u$ from
the measured Stokes parameters of KH 15D to correct for the instrumental
polarization.  
At Palomar the instrumental polarization was found to be $0.30\pm 0.22$\% 
over $4600-5600$ \AA\ and $0.13 \pm 0.13$\% over $5650-8500$ \AA.
At Keck the instrumental polarization was negligible, $< 0.1$\%.
At Palomar the observations were obtained through clouds, while the
Keck nights were mostly clear, but not photometric.

\begin{deluxetable}{ccc}
\tablewidth{0pt}
\tablehead{\colhead{Instrument}&\colhead{$\lambda$ range}&\colhead{Resolution} \cr
\colhead{\& side   } &\colhead{ (\AA)} &\colhead{(\AA/pixel)}}
\startdata
DBSP,  red  & 5620-8120 & 2.45 \cr
DBSP,  blue & 3350-5750 & 3.38 \cr
LRISp, red  & 3200-5850 & 2.0  \cr
LRISp, blue & 5650-9400 & 2.0  \cr
\enddata
\end{deluxetable}

\begin{deluxetable}{llccl}
\tablewidth{0pt}
\tablehead{\colhead{Date}&\colhead{Telescope}&\colhead{Time}&\colhead{Slit Width}& \colhead{Conditions}\cr
\colhead{(UT)}&\colhead{}&\colhead{(hr)}&\colhead{}&\colhead{}}
\startdata
2003 Jan 25 & Palomar & 3 &2$\farcs$0 & Partly cloudy\cr
2003 Feb 23 & Keck    & 3 &1$\farcs$5 & Partly cloudy\cr
2003 Feb 24 & Keck    & 1 &1$\farcs$0 & Clear\cr
2003 Feb 25 & Keck    & 2 &1$\farcs$5 & Clear\cr
\enddata
\end{deluxetable}

A 7th magnitude B2III star, HD 47887, is located 39$\arcsec$ away
from KH 15D.   In the Keck observations a diffraction spike
from this bright star periodically swept through the slit,
passing within several arcseconds of KH 15D in several exposures.  In
these exposures sky subtraction was difficult, making it impossible to 
accurately measure the polarization and also adversely affecting the 
color of the extracted spectra for two of the six hours.

Figure 1 depicts the phase of our observations with respect to the 48.36
day period of KH 15D.

\begin{figure*}[t]
\centerline{\psfig{file=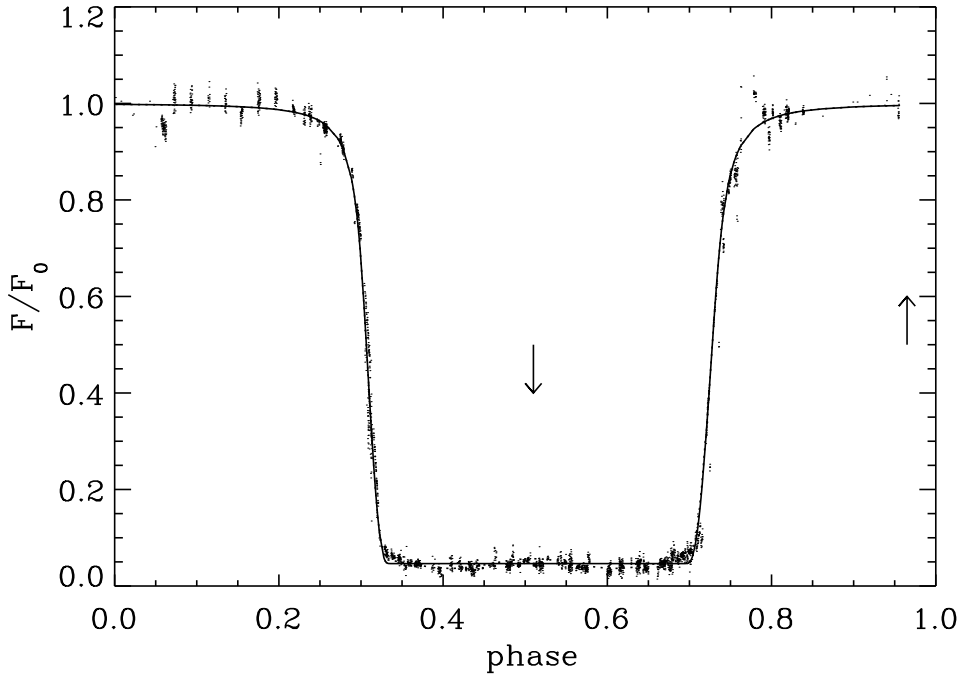,width=5in}} 
\figcaption{ Phased lightcurve of KH 15D from 2001-2002 campaign
(Herbst et al. 2002).  The arrows indicate the phases of our observations.
The solid line is a model fit to the data described in \S4.3.}
\label{lightcurve}
\end{figure*}

\section{Results}

The continuum polarization in the Keck and Palomar data was measured 
over wavelength bins of width 1000 \AA.
We averaged the 3 measurements from Palomar and subtracted 
off the mean polarization measured from observations of 4 null-standard stars.  
We computed the mean of 4 measurements from Keck and the error bars were computed 
from the standard deviation of these measurements plus the statistical error of the individual measurements.  
The Palomar data show a small enough polarization and large enough errors that 
the data are consistent with being unpolarized at the $\sim$ 1 $\sigma$ level
(see Figure 2).  However, the Keck data showed a much higher polarization, 
and are inconsistent with the Palomar data at a high significance.  
Tables 3 and 4 summarize the polarization measurements of the eclipsed and 
uneclipsed data.   The eclipsed data show a slight increase in polarization 
from the blue (4600-5600 \AA) to the red (7000-8000 \AA) at 3 $\sigma$ 
significance.  Most of this change is due to an increase in the $q$ Stokes
parameter (Figure 2), while $u$ remains constant; however, the errors are
large enough that a change in the polarization angle is only significant
at 1.5 $\sigma$.

\begin{deluxetable}{lccc}
\tablewidth{0pt}
\tablecaption{Palomar Observational Summary}
\tablehead{\colhead{Band (\AA)}&\colhead{$p$(\%)}&\colhead{$\theta_p$($^\circ$)}&
\colhead{$\langle F_{un}/F_{ec}\rangle$\tablenotemark{a}}}
\startdata
4600--5600   & 0.26$\pm$0.25& 31$\pm$27 & 1.00$\pm$0.04\cr
6000--7000   & 0.17$\pm$0.21&-19$\pm$35 &  1\cr
7000--8000   & 0.16$\pm$0.23&-22$\pm$41 & 1.05$\pm$0.04\cr
\enddata
\tablenotetext{a}{The mean of the ratio of the uneclipsed (Palomar) flux to
the eclipsed (Keck) flux within the given band, emission lines excluded,
divided by the ratio in the 6000--7000 band.}
\end{deluxetable}

\begin{deluxetable}{lcc}
\tablewidth{0pt}
\tablecaption{Keck Observational Summary}
\tablehead{\colhead{Band (\AA)}&\colhead{$p$(\%)}&\colhead{$\theta_p$($^\circ$)}}
\startdata
4000--4600   & 1.46$\pm$0.70& 23.9$\pm$14.\cr
4600--5600   & 1.63$\pm$0.19& 21.8$\pm$3.3\cr
6000--7000   & 1.85$\pm$0.14& 18.0$\pm$2.2\cr
7000--8000   & 2.25$\pm$0.10& 16.8$\pm$1.2\cr
8000--9000   & 2.24$\pm$0.29& 15.5$\pm$3.7\cr
\enddata
\end{deluxetable}

{\centerline{\psfig{file=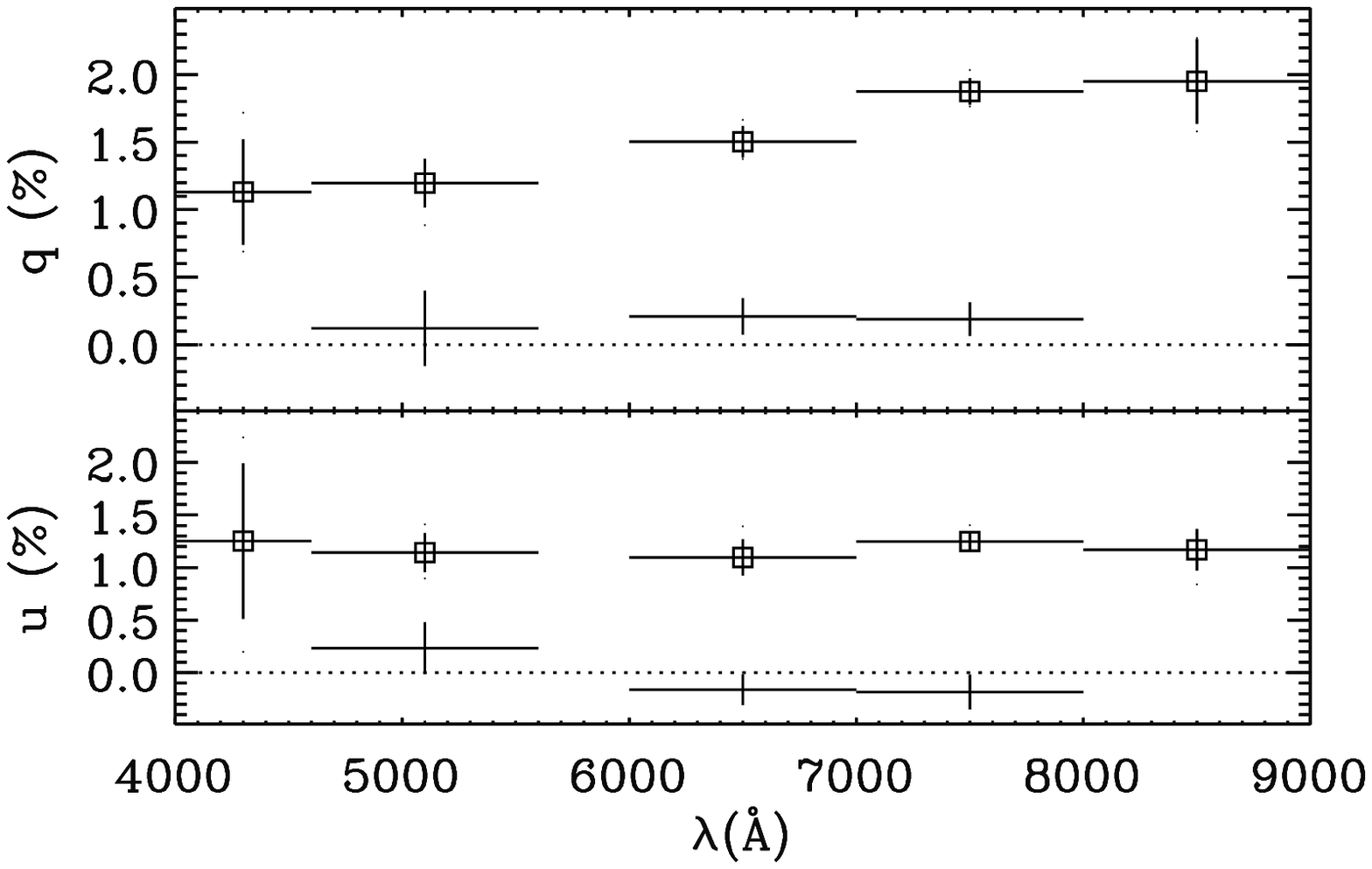,width=\hsize}} 
\figcaption{ Normalized Stokes parameters $q$ and $u$ of KH15D.
The squares are the data in eclipse, while the plus signs are data
out of eclipse. Horizontal bars represent the wavelength ranges over
which the Stokes parameters were measured.}}

Figure 3 shows the total flux spectrum in and out of eclipse.  The spectral
shape appears nearly identical in and out of eclipse, despite the change
in total flux of $\sim3.5$ magnitudes.  We measured the mean of the ratio of 
the uneclipsed to the
eclipsed data in each of the three bands (4600--5600, 6000-7000, and 
7000-8000 \AA), excluding the emission lines, and 
computed the standard deviation of the ratio within each band, which is
reported in Table 3.  The flux ratios differ between the bluest (4600-5600 \AA)
and reddest (7000-8000 \AA) bands by only 1 $\sigma$, which is statistically
consistent with no color change.  The absolute value of the uneclipsed to 
eclipsed flux ratios are not to be trusted as the nights were not photometric,
so we report only the ratio of each band to the 6000--7000 \AA\ band. 
For comparison with the photometric data of \citet{her03} we also computed
synthetic photometry of our spectra which gave $V-R=0.81$ mag and $V-I=1.45$ 
mag  during 
eclipse and $V-R=0.79$ mag outside of eclipse, which are in agreement with the range
of values reported by \citet{her03}.  We estimate the uncertainty on these colors 
to be $\sim 0.1$ mag as there appear to be differences between individual spectra 
on scales of $\sim 500$ \AA\ which may be due to fluctuations in the flat-fields, 
difficulty of sky subtraction, overall systematic uncertainties in flux calibration, or 
other instrumental variations. 

There are a large number of stellar photospheric
features which are seen in both spectra and are indicative of a K6V spectral type
with $T_{eff}\sim 4300$ K, somewhat hotter than the K7V type estimated by 
\citet{ham01} which may be due to the fact that our spectrum covers bluer 
wavelengths.  Almost all of these features show the same equivalent widths in 
both spectra, indicating that the eclipsed spectrum is a scaled version of the
uneclipsed spectrum.  We have measured the Lick indices \citep{wor97} for both 
the eclipsed and uneclipsed spectra, plotted in Figure 4.  The H$\delta$ indices
are contaminated by the [SII] emission line in the Keck data.  The Na D
equivalent width is smaller during the eclipse, which we attribute to unobscured 
Na D emission filling in the absorption line.  The 
measured indices for KH 15D agree best with those of a K6V star;  the
agreement is good enough that there is no evidence for optical veiling.
The H$\alpha$ emission line has a much higher flux
in the eclipsed spectrum than in the uneclipsed spectrum by a factor
of $\sim 3.3$ (since our data were not photometric, this factor is 
probably $\sim 2.7$).  This may be partly due to variability and partly
due to partial eclipse of the H$\alpha$ line as \citet{ham03} found that 
the flux of H$\alpha$ could decrease or increase during eclipse and
the shape of H$\alpha$ varied significantly as well.
They measured an equivalent width of 2 \AA\ outside eclipse and 40-70 \AA\
in eclipse, compared with our measurements of 3 \AA\ and 30 \AA, respectively.
In addition, four emission lines appear in the eclipsed spectrum that are not 
apparent in the uneclipsed spectrum.  The fluxes and equivalent widths of the 
emission lines are given in Table 5, which we have computed by subtracting off a
continuum template, with errors estimated from the standard deviation of the
residuals.  Included is the Na D absorption line at $\lambda 5890$ \AA\    
which is stronger in the uneclipsed spectrum than in the eclipsed spectrum
(we have used the uneclipsed spectrum as the continuum template for this line).

\begin{deluxetable}{lcc}
\tablewidth{0pt}
\tablecaption{Emission Lines}
\tablehead{\colhead{Line}&\colhead{Flux\tablenotemark{a}}&\colhead{EW}}
\startdata
\ \ \ \ \   (\AA)   &  ($10^{-15}$ erg cm$^{-2}$ s$^{-1})$ & (\AA)\cr
\hline
{\small H$\alpha$} $\lambda$ 6563\tablenotemark{b} & 5.88 $\pm$0.71   &  3.2$\pm$0.39\cr
\hline
{\small H$\alpha$} $\lambda$ 6563& 1.74 $\pm$0.19   & 29.7$\pm$3.2 \cr
[\ion{S}{2}] $\lambda$ 4069   & 0.362$\pm$0.025  & 22.2$\pm$1.5 \cr
{\small Na D} $\lambda$ 5890    & 0.153$\pm$0.015  &  3.6$\pm$0.36\cr
[\ion{S}{2}] $\lambda   
\lambda$ 6716,6731    & 0.093$\pm$0.013  &  1.6$\pm$0.23\cr
[\ion{O}{1}] $\lambda$ 6300    & 0.397$\pm$0.065  &  7.0$\pm$1.1 \cr
[\ion{O}{1}] $\lambda$ 6364    & 0.132$\pm$0.017  &  2.3$\pm$0.3 \cr
\enddata
\tablenotetext{a}{The absolute value of these measured fluxes is unreliable due
to non-photometric conditions and slit losses.}
\tablenotetext{b}{Uneclipsed line (the rest are eclipsed).}
\end{deluxetable}

\begin{figure*}[t]
\centerline{\psfig{file=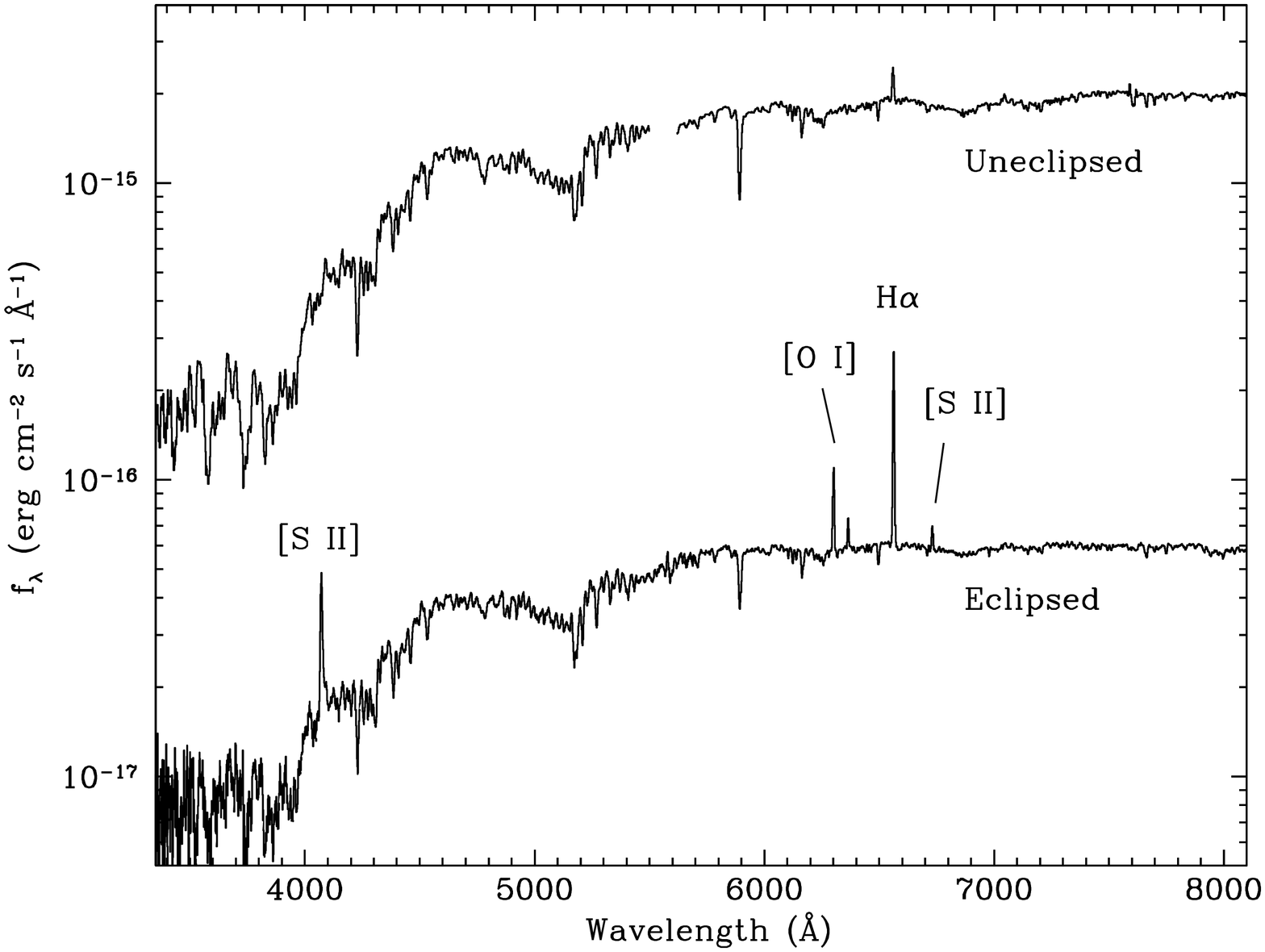,angle=-90,width=6in}} 
\figcaption{ Spectrum out of eclipse (top line) and in eclipse (bottom
line).}
\end{figure*}

\begin{figure*}[t]
\centerline{\psfig{file=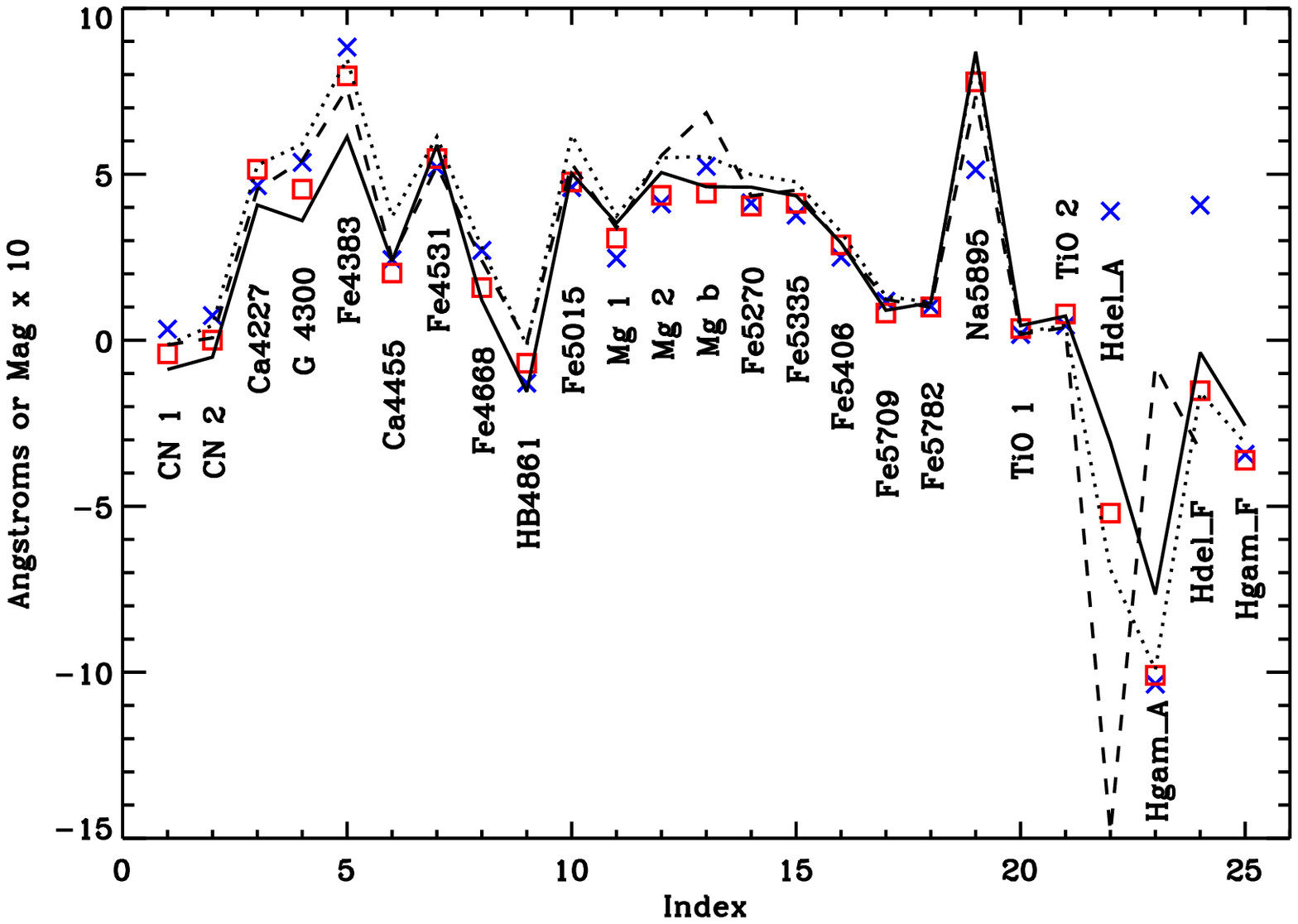,width=5in}} 
\figcaption{Lick spectral indices for the eclipsed (blue crosses) and
uneclipsed (red squares) data, as well as a K7V star (solid line),
K6V star (dotted line), and K5V star (dashed line). Note that the molecular
band indices with units of magnitudes (CN, Mg, and TiO) have been multiplied 
by a factor of 10 for clarity. The index numbers are taken from \citet{wor97}.}
\end{figure*}

\section{Discussion}

The variation of the polarization between eclipse and non-eclipse indicates
that the {\em scattering region is not completely obscured by the occulting material}.
If the occulter completely blocks all the light from the star, if the
scattering is nearly isotropic, and if the scattering region is nearly
spherically symmetric, then the residual flux, $\sim 5$\%, indicates that 
the optical depth of the unobscured scattering material is $\sim 0.05$.  
Since the spectra inside and outside eclipse are consistent with no change 
in the spectral shape, and since the spectrum within eclipse is consistent 
with very weak wavelength dependence of the polarization, we conclude that the 
scattering mechanism is nearly achromatic, implying that the scatterers may be 
electrons or large grains.  We discuss each of these possibilities in 
turn.

\subsection{Electron scattering} 

If the scattering mechanism were electrons, then the low measured polarization 
of 2\% indicates that the scattering region need be only slightly asymmetric.
For example, in the model of \citet{bro77}, the polarization
caused by an axisymmetric Thomson scattering envelope is
$P=\sin^2{i}/(2\alpha + \sin^2{i})$, where $\alpha$ is a ``shape
factor'' which depends on the precise geometry of the scatterers.
Since the eclipsing region is likely in the equatorial plane of
the circumstellar material, we can assume that $i \sim 90^\circ$.
Thus, $\alpha \sim 25$ for $P\sim 2$\%.  For a scattering geometry which
consists of a slightly oblate or prolate spheroid of axis ratio
$a=1+\epsilon$, then $\alpha \sim 5/\epsilon$, so $\vert \epsilon
\vert \sim 0.2$ to create the observed polarization, i.e. only a 20\%
deviation from spherical symmetry is required to give the measured
polarization.  Note that this
example is simply illustrative since we have no preferred model for
the scattering region.

If we assume that all of the light during eclipse is due to electron
scattering, then we can estimate the bound-free continuum emission expected 
from this gas.  Consider a nearly symmetric region of uniform electron
density and temperature surrounding the star of size $r_{es}$.
We assume that the eclipsing material is located near the
Keplerian radius, $r_{Kep}$, for a 48 day period.  The low polarization
implies that the electron scattering region must be at a larger radius, 
$r_{es}> r_{Kep} \sim 0.2$ AU to avoid becoming too asymmetric due to
obscuration.  Since the material is nearly 
symmetric, the mean optical depth of the scattering region is
$\tau_{es}\sim 0.05$, which implies an electron number density of 
$n_e < 2\times 10^{10}$ cm$^{-3}$ (assuming a uniform density).  The bound-free
emission expected at, say, 6000 \AA\ must be less than $F_\nu \sim 10^{-28}$
erg cm$^{-2}$ s$^{-1} Hz^{-1}$ to avoid changing the strength of the Lick
indices.  Computing the emission from an optically-thin spherical region of plasma in 
LTE, we find that $T > 2\times 10^6$ K to be consistent with
this upper limit.   Note that this is a strong lower limit since it
is likely that the scattering region need be larger to give the small
observed polarization.  In addition, one can place a constraint on
the temperature based on the strength of Balmer line emission; however,
the interpretation is complicated by the unknown optical depth of the Balmer
lines.  Even so, the lower limit on the temperature based on the lack
of detected bound-free emission is four times the virial temperature at
$r_{Kep}$, meaning that this gas would be strongly unbound and will 
outflow in less than a dynamical timescale, implying a mass loss rate of 
$10^{-7} M_\odot$ yr $^{-1}$, which is rather large for a weak-line T Tauri star.  
Thus, it is more likely that the scattering is not due to electrons, 
but may be due to large dust grains, which we discuss next.

\subsection{Dust scattering}

If the scattering were due to dust, then the grain sizes would have to
be fairly large to allow the weak wavelength dependence of the 
polarization {\em and} flux.  This is true for the zodiacal light which
reflects an optical spectrum that is identical to that of the Sun \citep{lei98}.
Based on the achromaticity of the zodiacal scattering, the scatterers are
thought to be large dust grains, 10-100 $\mu$m in size, and there is
further evidence that these grains might be quite ``fluffy'' (i.e.
low filling factor) with densities as low as $\rho_f \sim 0.25$ g cm$^{-3}$.

For the estimation of the wavelength dependence of the scattering
and polarization by dust particles we approximate the particles
as spheres (Mie scattering) and compute the polarized differential scattering 
cross section by single particles using a standard approach
\citep{vos02,wis80,boh83,wol00}.\footnote{{The code for calculation of the Mie scattering 
coefficients is available at\hfill\break}
{\tt http://mc.caltech.edu/$\sim$swolf/miex-web/miex.htm - or contact the authors.}}
We have chosen the distribution of refractive indices of the materials 
from \citet{wei01}:  62.5\% astronomical silicate and
37.5\% graphite.  Since graphite is crystalline, it was considered as two
``different'' materials in the computation: the first with a refractive index 
as measured perpendicular to the crystal rotation axis, 25\%, and the second 
parallel, 12.5\%.  We assumed that distribution of the dust size,
$a$, follows a power-law $dn/da \propto a^{-\alpha}$ for $a_{min}<a<a_{max}$,
and derived the scattering properties of the ensemble of particle
species and sizes \citep[][Wolf \& Voshchinnikov, in prep.]{mar78}.

We averaged the scattered polarized flux over the entire $4\pi$ steradians
divided by the total scattered flux to estimate the mean polarization as
a function of wavelength for each dust model.  We then compared this to
the relative wavelength dependence of the measured polarization (the
absolute polarization depends on the exact geometry of the scattering
region, which is unknown; see the next section for one particular
model).  We also compared the relative wavelength dependence of the
scattered flux by dividing the eclipsed spectrum by the uneclipsed spectrum.
Given the assumptions that (a) the wavelength dependence of the flux and
polarization does not depend on the geometry of the scatterers, and (b) the 
scatterers can be described by a single power-law, we can then place a
constraint on the parameters of the dust model.  We find that the
lower limit, $a_{min}$, does not change the fit much, so we simply
vary $1\mu$m$<a_{max}<100\mu$m and $0<\alpha<2$.  The minimum 
$\chi^2$ is 12 for 5 data points for $a_{max}=8\mu$m and $\alpha=0$, with
the confidence contours plotted in Figure 5.  At 90\% confidence,
$\alpha <2$ and $5 \mu$m$< a_{max} < 10\mu$m, which indicates that
the scattering is dominated by large dust grains.  Unfortunately, the
minimum reduced $\chi^2_r= 2.2$,  which is rather large.  This is due
to the fact that the measured polarization increases toward longer
wavelengths, while the dust scattering models predict a polarization
that is nearly constant, or slightly decreasing, with wavelength.

\centerline{\psfig{file=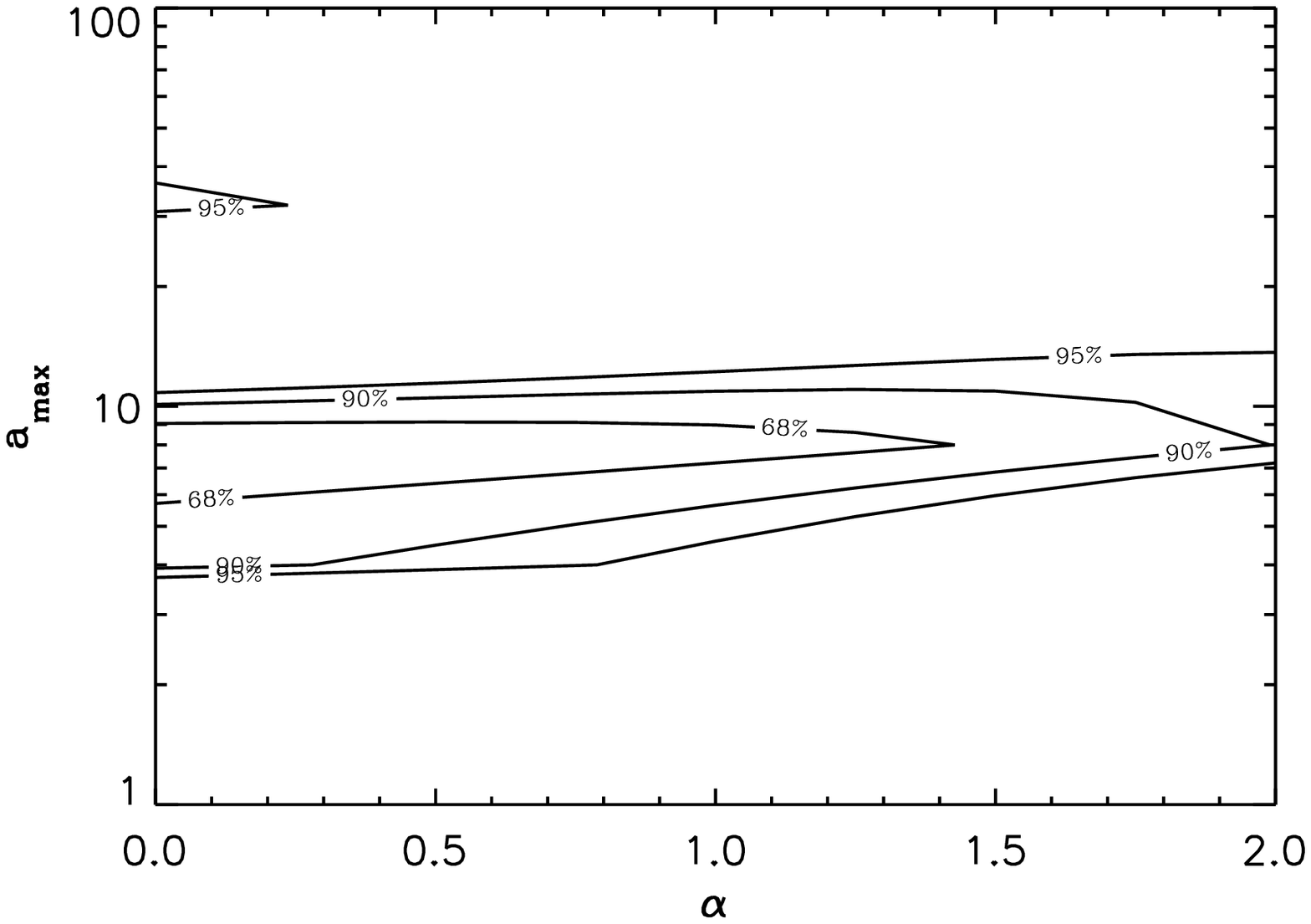,width=\hsize}} 
\figcaption{Confidence limits for dust models as a function of $a_{max}$
and $\alpha$.}

The opacity of large dust grains scales as $\kappa/\rho =0.75 a^{-1} 
\rho^{-1} \sim 3\times 10^3 a_{10}^{-1}(\rho_f/\rho)$ cm$^2$ g$^{-1}$, where 
$a_{10}=a/10 \mu$m. 
Thus, the Eddington limit for the smaller of the large dust grains
scales as $L_{\mathrm{Edd,dust}}=1.7\times 10^{34} a_{10} (\rho/\rho_f)$ erg s$^{-1}$, 
which is similar to the luminosity of KH 15D, $L_{\mathrm{KH 15D}} \sim 2.4
\times 10^{33}$ erg s$^{-1}$.
Thus, smaller dust grains will be removed from the system by radiation
forces.  In addition, radiation forces on the larger grains may be
significant enough to modify the structure of the system.

\subsection{Scattering Model}

\citet{her03} have proposed a model for the occulter which consists of
a semi-infinite edge crossing the star.  This model is attractive as the
finite size of the star and limb darkening can account for the shape of
the ingress and egress, and, if the occulter is at the Keplerian radius, 
then it must be inclined to the orbital plane by 15$^\circ$ to give
the correct duration of ingress and egress.  However, we have found that
the ingress and egress of the eclipse (Figure 1, near phase of 0.3 and
0.7) show a more gradual rolloff than expected for a semi-infinite occulter.
This can be fit with an optical depth of the occulter which scales
with position as $\tau \propto x^{-2}$, where $x$ is the coordinate
perpendicular to the edge of the occulter, which gives an excellent fit
to the phased lightcurve data (solid line in Figure 1).

This material causing extinction will scatter as well, and in the case of large
grains there will be significant forward scattering \citep{van81}.
In fact, this may account for the scattered flux present during the 
eclipse; however, to compute the scattered light requires a knowledge of
the three-dimensional distribution of scattering and occulting material,
as well as a knowledge of the scattering properties.  We construct such
a model by assuming that (a) the occulter is located at the Keplerian radius 
and (b) the 15$^\circ$ tilt of the occulter is due
to a warped disk.   We model the material as a succession of tilted circular
annuli, creating a warped disk, with the tilt angle given by
\begin{equation}
i_{tilt}=i_w \exp{\left[-{(r-r_w)^2\over 2\sigma_w^2}\right]},
\end{equation} 
where $r_w$ is the radius of maximum 
tilt, $i_w$ is the maximum tilt, $r$ is the radius in the disk, and $\sigma_w$ 
is the width of the tilted region, and we assume that the nodes of the warped 
annuli are aligned.  The observer we place at an angle $i_{obs}$ to the equatorial 
plane, and we rotate the warped disk as a solid body with a period equal to 
that of KH 15D.  In addition, we cut the disk off at $r_{in}=0.5r_w$ and 
$r_{out}=4r_w$.  The opacity of the scattering material we distribute as 
$\kappa=\kappa_0 [h(r/r_w)/(\theta-\pi+i_{tilt})]^2$,
to produce the observed roll-off in intensity,
where $h$ is the angular scale height, $\theta$ is the polar angle, and 
$\kappa_0$ is a fiducial opacity.
We find a good fit to the lightcurve for $i_w=17.6^{\circ}$, 
$i_{obs}=6.7^{\circ}$, $h=1.6^{\circ}$, $r_w=0.22$ AU (this is the Keplerian
radius), $\sigma_w=r_w$.  When taking into account only extinction, this
distribution produces the solid lightcurve in Figure 1 after the addition of 
4.5\% of the uneclipsed flux during eclipse.  However, part of the
extinction is due to scattering, so we also have computed the singly-scattered
component of this density distribution assuming an albedo of unity and using the
scattering function for zodiacal light derived by \citet{hon85} which 
is strongly forward scattering and has a maximum polarization of 22\%.  We 
find that the polarized flux of the scattered light during eclipse is 0.1\%
of the unobscured total flux, 
which is the same as the observed polarized flux during eclipse of KH 15D
(Figure 6).   Figure 7 shows the appearance of the scattered light in this
model at mid-eclipse (an animated version of this figure which shows the 
appearance as a function of phase is available on-line).

\begin{figure*}[t]
\centerline{\psfig{file=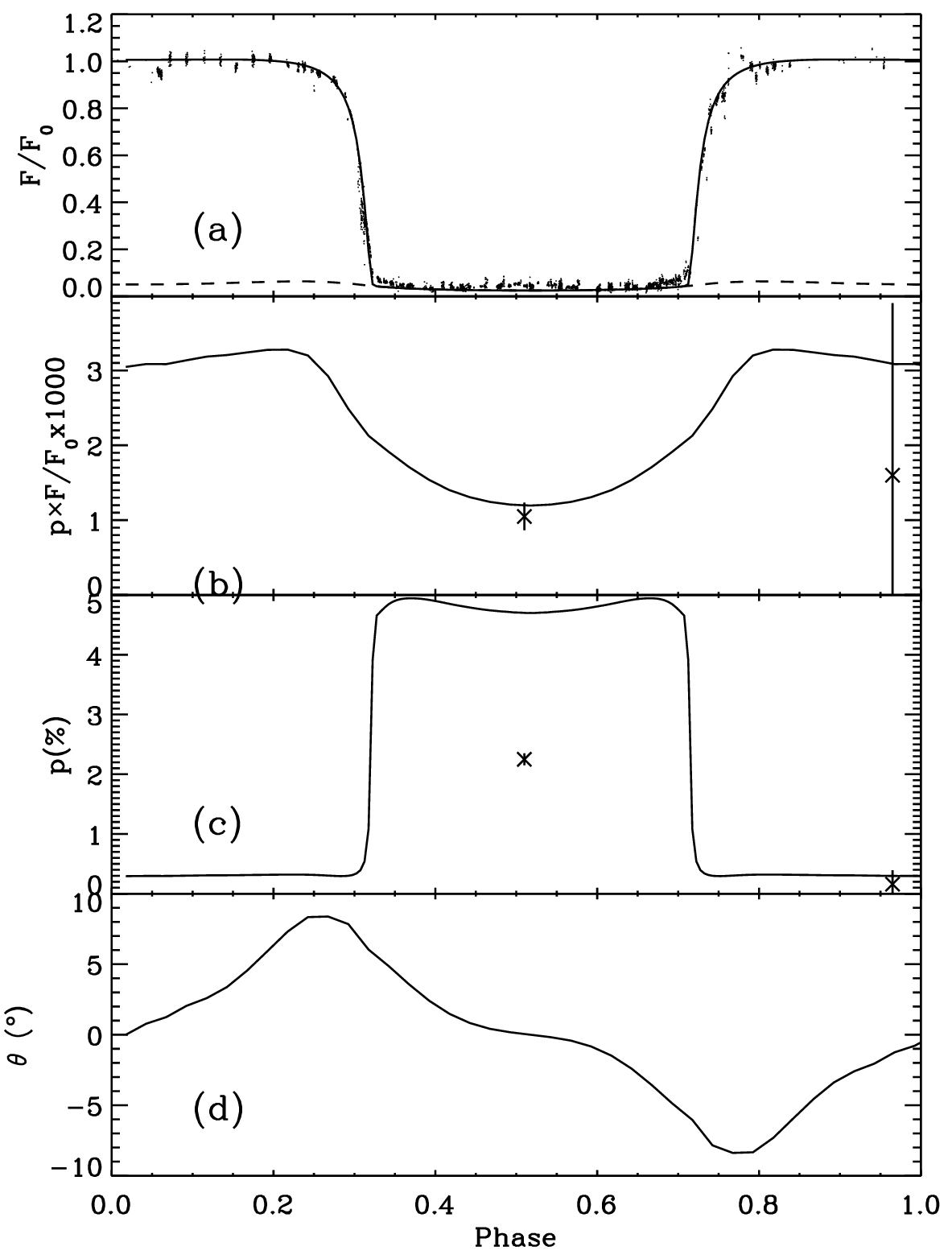,width=6.5in}} 
\figcaption{(a) Flux of KH 15D as a function of phase from \citet{her03}, along
with our model (solid curve) and the singly-scattered light (dashed curve).
(b) Model polarized flux normalized by $F_0$ at 7000-8000 \AA\ (unity would be 100\% polarized light).
At mid-eclipse, we have taken our observed polarization and multiplied by the
mean flux of the I-band observations near mid-eclipse by \citet{her03}, $F/F_0=0.047\pm0.01$,
since our observations were not photometric.
The crosses with error bars show our observations, which agree remarkably well
with the model. (c) Model polarization. Note the disagreement
at mid-eclipse which may be due to an unpolarized component which also produces
the disagreement in the total flux at mid-eclipse. (d) Polarization angle.  We
have not plotted the observed angles since we do not know the plane of the
disk on the sky and our temporal coverage is not sufficient to determine the shape
of $\theta(t)$.}
\end{figure*}

One prediction of this model is that the polarization angle
should vary by $\sim 15^\circ$ during the eclipse.  However, 
the total model flux during mid-eclipse is only 2.5\% of the uneclipsed flux
which is about half the observed mid-eclipse flux,
indicating that there may be another component of scattered 
light which has low polarization, such as diffracted light, which is not taken 
into account in the model of \citet{hon85}, or back-scattered light.
In a future study we may carry out more sophisticated 
modeling, varying the radius of the warp as well as the symmetry of the warp,
and including multiple scattering and varying the dust grain model, to see
how the scattered flux varies with dust composition and with
wavelength of the observed light, and we will see whether diffraction,
multiple-scattering, or back-scattering can explain the brightening at mid 
eclipse.  There is evidence that
the ingress and egress differ slightly in time between the even and odd eclipses,
which may imply that the warp is an $m=2$ mode rather than $m=1$ as in our model.  If
so, the back warp will be partly eclipsed by the front warp, while the edges will
be visible during mid-eclipse, which may explain the rebrightening.  Alternatively,
the symmetry during mid-eclipse of the front warp may cause a brightening due
to diffraction, which is likely to be wavelength dependent.  If the albedo is
less than unity, then mid-infrared observations may point toward an $m=2$ 
configuration if they also show a peak at mid-eclipse since the infrared light is 
emitted isotropically, similar to the back-scattered light.

\centerline{\psfig{file=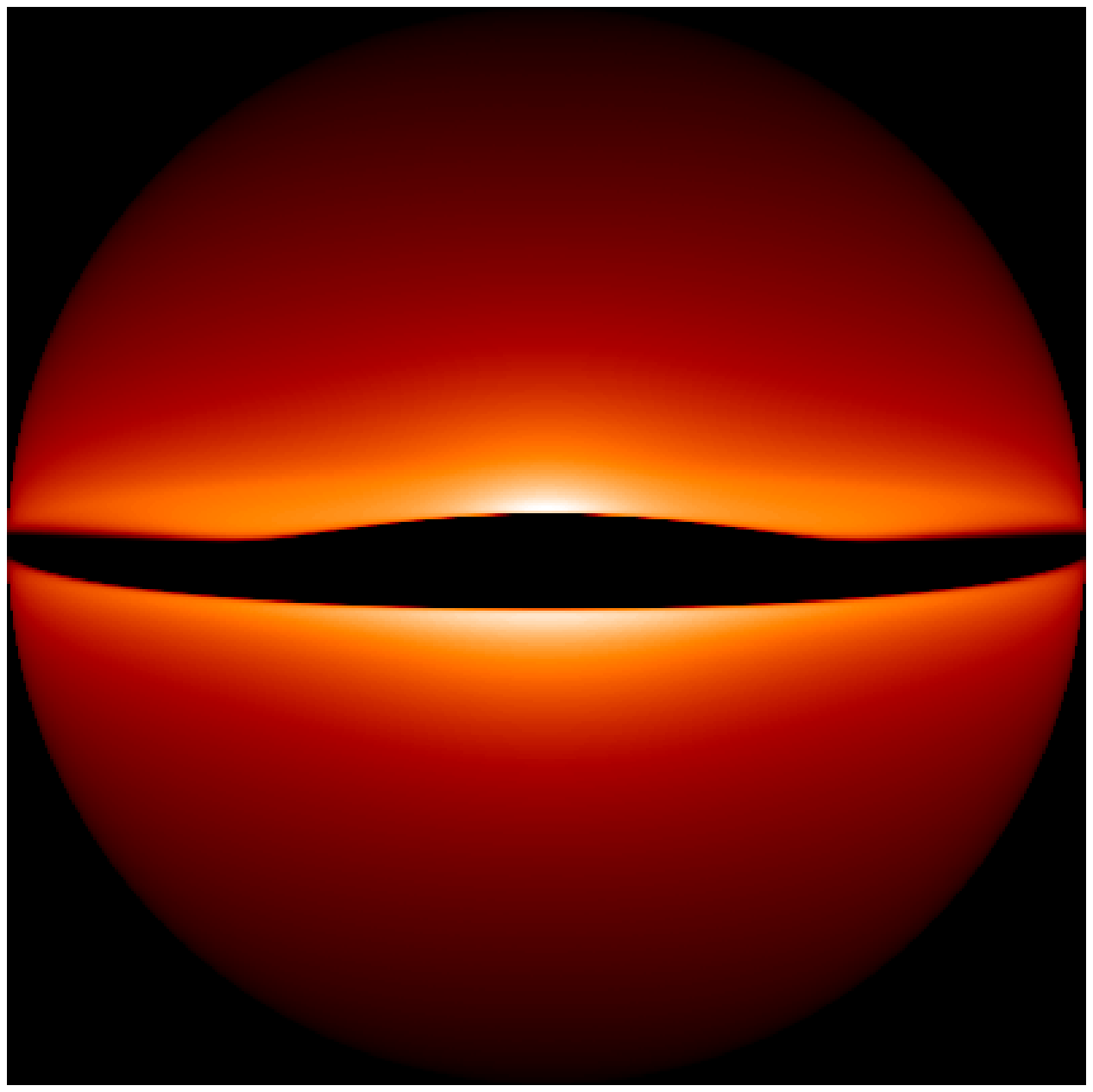,width=\hsize}} 
\figcaption{Scattered light at mid-eclipse for the model described in the text.}

\section{Conclusions}

A summary  of the results:

1)  Near mid-eclipse the polarization is $\sim$2\%, and is consistent with 
zero outside of eclipse (0.2$\pm$0.2\%).  Combined with other photometric 
data, the polarized flux near mid-eclipse is $\sim 0.1$\% of the flux outside
of eclipse.  This indicates that there is
an unobscured scattering region surrounding the T Tauri star which creates
the polarization; the polarization cannot be solely due to foreground 
transmission effects (not associated with the star) as there would no variation.

2)  The data are nearly consistent with no wavelength dependence of the 
polarization during eclipse.  This indicates that the scatterers are 
approximately grey.  In addition, electron scattering is unlikely 
to cause the observed polarization.

3)  The spectrum within the eclipse is consistent with being a scaled version
of the spectrum outside of eclipse with no wavelength dependence of the ratio 
except for the presence of stronger emission lines which also argues for grey
scattering;  Rayleigh scattering is ruled out.  Our Mie dust scattering model
implies $\alpha < 2$ and $5 \mu$m$< a_{max} < 10\mu$m at 90\% confidence;
however, the best fit model is not an adequate fit.

4)  The qualitative features of the eclipse can be reproduced by a model
of a warped disk that orbits in front of the star, periodically eclipsing
the star, having an extended atmosphere which causes a decrease in light
before and after eclipse as well as scattered light within eclipse, which
is consistent with the observed polarized flux.

The polarization adds an extra clue that the eclipsing material completely
occults the star since the maximum observed polarization of classical
T Tauri stars is $\sim 2.5$\%  \citep{men91} and the KH 15D spectrum looks like
that of a classical T Tauri star during eclipse.  This adds to the evidence that 
the eclipse is complete based on (a) the steep egress and ingress; (b) the flat 
bottom of the trough of the eclipse; (c) the fine tuning that would be required 
to arrange 95\% covering.  It is interesting that the trough of the eclipse does 
show some slight variations, namely an increase near mid-eclipse, which may 
indicate that the scattering
region is partially eclipsed and this eclipse varies with time.  It would be
interesting to measure the polarization as a function of time within the
eclipse to look for variation.

We have modeled the eclipse with a warped disk since a discrete cloud would be 
sheared in ten orbits, while the eclipses have repeated with similar
durations for $\sim$ 50 orbits.  However, this does not mean that the
warped disk model is without its own problems: If the warped model is correct, 
what drives the warp at such a short period?  The interaction of the stellar
magnetosphere is proposed to cause a similar occulting behavior in the
T Tauri star AA Tauri \citep{bou99};  however, this star has a much short period 
between occultations, 8 days, while the longer period of KH 15D is much longer than
the spin period of any known T Tauri star.
One can place a lower limit on the warp radius of $r_{min}=10 R_*$ since inwards of
this radius the warp would have to be at $>45^\circ$ to create the correct duration 
of ingress due to the finite size of the star.
At $r_{min}$, the Keplerian period is 0.8 days, while a warp precession timescale
is typically much longer than the Keplerian timescale.  It may be that a
planet excites a wave in the disk \citep{win03} or that an elongated dusty
vortex causes obscuration \citep{bar03};  each of these models might be consistent
with the observed polarization profile, which remains to be quantified with a
scattering calculation similar to the one presented here.

Future observations of interest would be to (1) measure the magnitude and angle
of polarization throughout ingress and/or egress to test the warp model;
(2) measure the polarization at higher spectral resolution within the H$\alpha$ 
emission line to see whether it is unpolarized;  (3) look for variations of the 
flux during eclipse to determine whether the stellar variability is washed out 
by the time delays from scattering (although this requires a rather extended 
scattering region).  
In addition, near- and mid-infrared observations may show fluctuations if the occulting
region has a warped geometry, as we have suggested.

\acknowledgments

We thank Catrina Hamilton, Lynne Hillenbrand, Patrick Ogle, and Russel White for 
useful discussions and Marshall Cohen for exchange of observing time.  We thank
Bill Herbst for providing published photometric data.  Some of the data presented 
herein were obtained at the W.M. Keck Observatory, which is operated as a scientific 
partnership among the California Institute of Technology, the University of 
California and the National Aeronautics and Space Administration. The 
Observatory was made possible by the generous financial support of the W.M. 
Keck Foundation. The authors wish to recognize and acknowledge the very 
significant cultural role and reverence that the summit of Mauna Kea has 
always had within the indigenous Hawaiian community.  We are most fortunate to 
have the opportunity to conduct observations from this mountain.  

Support for E. A. was provided by the National Aeronautics and Space Administration
through Chandra Postdoctoral Fellowship Award PF0-10013 issued by the Chandra
X-ray Observatory Center, which is operated by the Smithsonian Astrophysical
Observatory for and on behalf of the National Aeronautics Space Administration
under contract NAS 8-39073.  S. Wolf was supported by NASA through grant
NAG~5-11465.

Research by A. J. B. is supported by NASA through Hubble Fellowship grant 
HST-HF-01134.01-A awarded by STScI, which is operated by AURA, Inc., for NASA, 
under contract NAS 5-26555.


\begin{thebibliography}{}
\bibitem[Antonucci (1993)]{ant93} Antonucci, R., 1993, ARA\&A, 31, 473
\bibitem[Barge \& Viton(2003)]{bar03} Barge, P. \& Viton, M., 2003, ApJ, 593, L117
\bibitem[Bohren \& Huffman(1983)]{boh83} Bohren C.F., Huffman D.R., 1983,
  {\it Absorption and scattering of light by small particles,}
  John Wiley \& Sons, New York
\bibitem[Bouvier et al.(1999)]{bou99} Bouvier, J. et al., 1999, A\&A, 349, 619
\bibitem[Breger \& Dyck(1972)]{bre72} Breger, M. \& Dyck, H. M., 1972, ApJ, 175, 127
\bibitem[Brown \& McLean(1977)]{bro77}  Brown, J. C.; McLean, I. S., 1977, A\&A, 57, 141
\bibitem[Bryden et al.(2000)]   {bry00} Bryden, G., R${\rm {\grave o} {\dot z}}$yczka, M., 
Lin, D. N. C. \& Bodenheimer, P., 2000, ApJ, 540, 1091
\bibitem[Goodrich(1991)]{goo91} Goodrich, R., 1991, PASP, 103, 1314
\bibitem[Hamilton et al.(2001)] {ham01}Hamilton, C. M., Herbst, W., Shih, C. \& Ferro, A. J., 2001, ApJ, 554, L201
\bibitem[Hamilton et al.(2003)] {ham03}Hamilton, C. M., Herbst, W., Mundt, R., 
 Bailer-Jones, C. A. L., Johns-Krull, C. M., 2003, astro-ph/0305477
\bibitem[Herbst et al.(2003)]   {her03}Herbst, W. et al., 2003, PASP, 114, 1167
\bibitem[Hong(1985)] {hon85} Hong, S. S., 1985, A\&A, 146, 67
\bibitem[Kartje(1995)]{kar95} Kartje, J., 1995, ApJ, 452, 565
\bibitem[Kearns \& Herbst(1998)]{kea98} Kearns, K. E. \& Herbst, W., 1998, ApJ, 116, 261
\bibitem[Leinert et al.(1998)]{lei98} Leinert, Ch. et al., 1998, A\&AS, 127, 1
\bibitem[Martin(1978)]{mar78} Martin P.G., 1978, Cosmic dust.
 Its impact on astronomy, Clarendon Press, Oxford
\bibitem[M\'enard(1991)]{men91}M\'enard, F., 1991, in {\it IAU Colloq. 129}, 
ed. Bertout C., Collin-Souffrin, S., \& Lasota, J. P.  (Gif-sur-Yvette: 
Editions Frontieres), p. 59
\bibitem[M\'enard et al.(2003)]{men03}M\'enard, F., Bouvier, J., Dougados, C.,
 Melnikov, S. Y., \& Konstantin, N.  G., 2003, A\&A, in press, astro-ph/0306552
\bibitem[Miller et al.(1988)]{mil88} Miller, J. S., Robinson, L. B. \& 
Goodrich, R. W., 1988, in {\em Instrumentation for Ground-Based Optical
Astronomy}, ed. L. B. Robinson (New York: Springer), p. 157
\bibitem[Mekkaden(1999)]        {mek99}Mekkaden, M. V., 1999, A\&A, 344, 111
\bibitem[Ogle et al.(1999)]{ogl99} Ogle, P. M., Cohen, M. H., Miller, J. S., Tran, H. D., 
Goodrich, R. W. \& Martel, A. R., 1999, ApJS, 125, 1O
\bibitem[Oke et al.(1995)]      {oke95} Oke, J. B. et al., 1995, PASP, 107, 375
\bibitem[Oke \& Gunn(1982)]     {oke82} Oke, J. B. \& Gunn, J. E., 1982, PASP, 94, 586
\bibitem[Prugniel \& Soubiran(2001)]{pru01} Prugniel, Ph. \& Soubiran, C., 
2001, A\&A, 369, 1048
\bibitem[van de Hulst(1981)]{van81} van de Hulst, H. C., 1981, {\it Light Scattering by
Small Particles}, Dover: New York
\bibitem[Voshchinnikov(2002)]{vos02} Voshchinnikov N.V., 2002, Astrophys. Space 
Phys. Rev., 12, 1
\bibitem[Weingartner \& Draine(2001)]{wei01} Weingartner, J. C. \& Draine, B. T., 
2001, ApJ, 548, 296
\bibitem[Winn et al.(2003)]{win03}Winn, J. N., Garnavich, P. M., Stanek, K. Z., 
Sasselov, D. D., 2003, ApJ, 593, L121
\bibitem[Wiscombe(1980)]{wis80} Wiscombe W.J., 1980, {\rm Appl. Optics}, 19, 1505
\bibitem[Wolf(2000)] {wol00}Wolf, S., 2000, Computer Physics Communications, 150, 99
\bibitem[Wolf et al.(1999)]     {wol99}Wolf, S.,  Henning, Th. \& Stecklum B., 1999, A\&A, 349, 839
\bibitem[Worthey \& Ottaviani(1997)]{wor97} Worthey, G. \& Ottaviani, D. L., 1997, ApJS, 111, 377
\end{thebibliography}
\end{document}